\documentclass[twocolumn]{aastex63}

\submitjournal{AAS Journals on Dec 15, 2021}

\graphicspath{{./}{figures/}}

\begin{document}

\title{ALMA measures molecular gas reservoirs comparable to field galaxies in a low-mass galaxy cluster at $z=1.3$}

\author[0000-0003-2919-7495]{Christina C. Williams}
\affiliation{Steward Observatory, University of Arizona, 933 North Cherry Avenue, Tucson, AZ 85721, USA}

\author[0000-0002-8909-8782]{Stacey Alberts}
\affiliation{Steward Observatory, University of Arizona, 933 North Cherry Avenue, Tucson, AZ 85721, USA}

\author[0000-0003-3256-5615]{Justin Spilker}
\altaffiliation{NHFP Hubble Fellow}
\affiliation{Department of Astronomy, University of Texas at Austin, 2515 Speedway, Stop C1400, Austin, TX 78712, USA}
\affiliation{Department of Physics and Astronomy and George P. and Cynthia Woods Mitchell Institute for Fundamental Physics and Astronomy, Texas A\&M University, 4242 TAMU, College Station, TX 77843-4242}

\author[0000-0003-1832-4137]{Allison G. Noble}
\affiliation{ School of Earth and Space Exploration, Arizona State University, Tempe, AZ 85287-1404,USA}

\author[0000-0001-7768-5309]{Mauro Stefanon}
\affiliation{Leiden Observatory, Leiden University, NL-2300 RA Leiden,Netherlands}

\author[0000-0001-9262-9997]{Christopher N. A. Willmer}
\affiliation{Steward Observatory, University of Arizona, 933 North Cherry Avenue, Tucson, AZ 85721, USA}

\author[0000-0001-5063-8254]{Rachel Bezanson}
\affiliation{Department of Physics and Astronomy and PITT PACC, University of Pittsburgh, Pittsburgh, PA, 15260, USA}

\author[0000-0002-7064-4309]{Desika Narayanan}
\affiliation{Department of Astronomy, University of Florida, 211 Bryant Space Science Center, Gainesville, FL 32611, USA}
\affiliation{Cosmic Dawn Center (DAWN)}

\author[0000-0001-7160-3632]{Katherine E. Whitaker}
\affiliation{Department of Astronomy, University of Massachusetts, Amherst, 710 N. Pleasant Street, Amherst, MA 01003, USA}
\affiliation{Cosmic Dawn Center (DAWN)}

\begin{abstract}
We report the serendipitous discovery of an overdensity of CO emitters in an X-ray-identified cluster (Log$_{10}$M$_{\rm halo}/M_{\odot}\sim13.6$  at $z=1.3188$) using ALMA. We present spectroscopic confirmation of 6 new cluster members exhibiting CO(2-1) emission, adding to 2 existing optical/IR spectroscopic members undetected in CO.
This is the lowest mass cluster to date at $z>1$ with molecular gas measurements, bridging the observational gap between galaxies in the more extreme, well-studied clusters (Log$_{10}$~M$_{\rm halo}/M_{\odot}\gtrsim14$) and those in group or field environments at cosmic noon.
The CO sources are concentrated on the sky (within $\sim$1--arcmin diameter) and phase space analysis indicates the gas resides in galaxies already within the cluster environment. 
We find that CO sources sit in similar phase space as CO-rich galaxies in more massive clusters at similar redshifts (have similar accretion histories) while maintaining field-like molecular gas reservoirs, compared to scaling relations.
This work presents the deepest CO survey to date in a galaxy cluster at $z>1$, uncovering gas reservoirs down to M$_{\rm H_{2}}>1.6\times10^{10}$M$_{\odot}$ (5$\sigma$ at 50\% primary beam).
Our deep limits rule out the presence of gas content in excess of the field scaling relations; however, combined with literature CO detections, cluster gas fractions in general appear systematically high, on the upper envelope or above the field.
This study is the first demonstration that low mass clusters at $z\sim1-2$ can host overdensities of CO emitters with surviving gas reservoirs, in line with the prediction that quenching is delayed after first infall while galaxies consume the gas bound to the disk. 
\end{abstract}

 
\section{Introduction} \label{sec:intro}

One of the most transformational events in the lives of galaxies is the cessation of active star-formation (quenching), marking a transition to passive evolution. Environment is a strong regulator of star formation activity, operating independently of mass-dependent quenching to $z<1$, 
and producing largely quenched populations in the low redshift Universe \citep{Dressler1980, Balogh1998, Lewis2002, Peng2010}. 

By the present day ($z=0$) the era of active galaxy growth in clusters through star formation has mostly completed. Moreover, at $z<1$, clusters are already largely quiescent in their cores \citep[e.g.][]{Muzzin2012, Patel2009, Finn2010, Vulcani2010}. However, at z=1-2,  discovery of significant populations of starforming galaxies in clusters \citep[i.e.][]{Cooper2006, Hilton2010, Tran2010, Fassbender2011, Fassbender2014, Hayashi2011, Tadaki2011, Brodwin2013, Zeimann2013, Alberts2014, Bayliss2014, Santos2014, Santos2015, Ma2015, Alberts2016, Alberts2021} indicate a reversal in the star formation rate (SFR)-density relation at higher redshifts, albeit with significant cluster-to-cluster variation \citep[i.e.][]{Alberts2016}.
Massive clusters at this transition epoch from $z=1-2$ have been shown to host field-like (obscured) star formation \citep{Brodwin2013,Alberts2014,Alberts2016, Alberts2021}, alongside a significant ramp up of the environmental quenching efficiency \citep{Nantais2017}. Understanding this epoch is pivotal to our understanding of environment over cosmic time, linking the proto-cluster regime to local clusters through a critical era of galaxy buildup.

Finally, at $z>2$, the majority of significant overdensities are expected to be proto-clusters, structures that are not yet virialized, lacking an established hot intra-cluster medium (ICM).  Proto-clusters are typically signposted by their dominant population of star forming galaxies (traced by Lyman-$\alpha$, H-$\alpha$, or dust continuum emission). Proto-clusters likely contribute a significant portion of the cosmic star formation rate density, while  in this active phase of growth via star formation \citep{Chiang2017}, prior to the era of increased quenching efficiency at $z=1-2$. 

Quenching processes likely initiate during infall into the dense cluster environment. 
A critical constraint on the impact of environment is the cold molecular gas content in galaxies, the fuel for star formation. Unfortunately, the number of true virialized galaxy cluster candidates (confirmed by a hot ICM, or inferred from a red sequence) at this critical era of $1<z<2$ that also have molecular gas constraints is small \citep{Noble2017,Hayashi2017,Rudnick2017,Stach2017,Coogan2018,Tadaki2019}. This is a severe limitation given that current studies indicate significant cluster-to-cluster variation in star formation activity \citep{Geach2006, Alberts2016}. 
Targeted CO observations in clusters tend to be conservative, choosing to observe rich overdensities to maximize success rate. The known CO emitters in these high-mass clusters to date exhibit field-like or enhanced molecular gas content \citep[i.e.][]{Noble2017,Rudnick2017,Hayashi2018}, suggesting that enhanced gas fractions may be prevalent in high density environments at high redshift. However, due to the challenging nature of high-redshift CO observations, these surveys have relatively high molecular gas detection limits, which can cause a biased picture by detecting only the most gas rich sources.
Blind and deep CO spectroscopy would provide a less biased and more comprehensive picture of gas content in cluster galaxies, but unfortunately is very expensive to obtain.

Fortuitously, we recently conducted an ultra-deep Atacama Large Millimeter/submillimeter Array (ALMA) spectroscopic campaign targeting CO(2-1) emission in massive quiescent galaxies at z$\sim$1.5 \citep{Williams2021}. Our data serendipitiously discovered that one target resides in a previously unknown overdensity of CO emitters. The structure is an X-ray identified galaxy cluster with a halo mass of Log$_{10}$M$_{\rm halo}$/M$_{\odot}=13.6$  \citep{Goz2019} that, until our ALMA discovery, lacked spectroscopic confirmation. We have identified 6 cluster members based on their CO(2-1) emission within 225 kpc of the central quiescent galaxy. 
This low mass cluster presents a rare addition to the current CO-observed samples, expanding the halo mass range covered at $z>1$, and thus represents an opportunity to study previously unexplored parameter space in cluster evolution.

\begin{figure*}[th]
    \includegraphics[scale=.35, trim=50 0 70 0, clip]{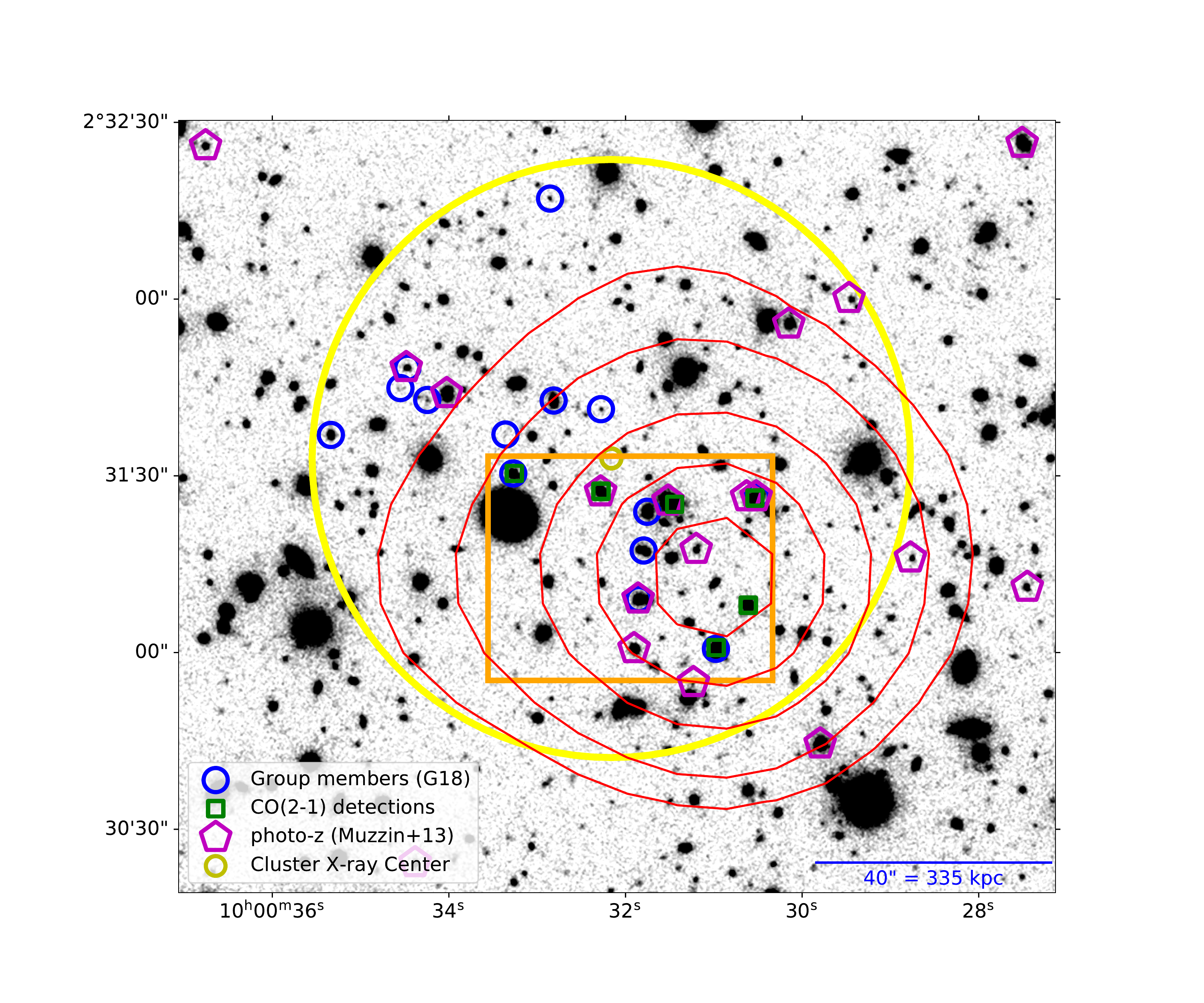}
    \includegraphics[scale=.43, trim=100 10 20 10, clip]{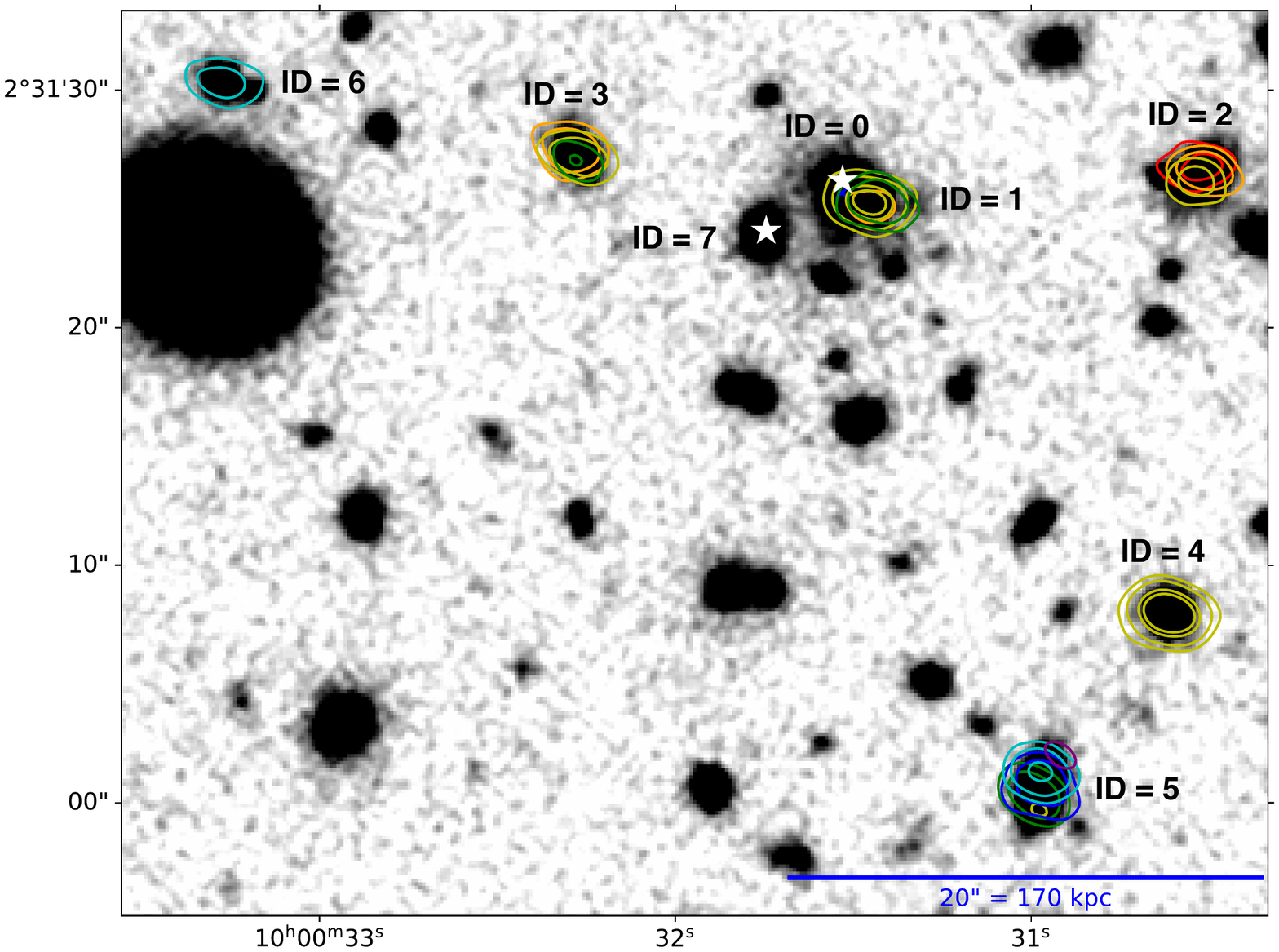}
    \caption{UltraVISTA K-band image of the cluster field (grayscale) identifying cluster members. Left: large scale image of the galaxy cluster with Voronoi galaxy overdensity per comoving Mpc$^{-2}$, in a redshift slice $1.28<z<1.34$ based on photometric redshift \citep[red contours;][]{Scoville2013}. Galaxy group members \citep{Goz2019} are in blue and other galaxies with similar photometric redshift of $1.28<z<1.35$ in magenta.  New CO(2-1) detections from this work are shown in green. The X-ray cluster center is indicated by a small yellow circle with R$_{200}$ as larger yellow circle. Right: zoom-in on the region of the cluster with CO emitters. Contours indicate the CO(2-1) emission measured with ALMA (colors represent different 200 km/s velocity channels). Two spectroscopic members with no CO(2-1) emission are identified with white stars. }
    \label{fig:image}
\end{figure*}

In Section \ref{sec:data} we present the multiwavelength evidence for the galaxy cluster hosting our galaxies, our ALMA data and galaxy sample properties. In Sections \ref{sec:results} and \ref{sec:discussion} we present the cold molecular gas reservoir measurements and discuss our results in the context of both gas reservoirs measured in field galaxies, as well as other (typically more massive) clusters, offering a unique opportunity to bridge the observational gap between these vastly different galaxy environments at $z>1$.  Throughout this work we assume a $\Lambda$CDM cosmology with  H$_0$=70 km s$^{-1}$ Mpc$^{-1}$, $\Omega_M$ = 0.3, $\Omega_\Lambda$ = 0.7, and a \citet{Kroupa2001} initial mass function (IMF).

\section{Data} \label{sec:data}
\subsection{Evidence for the galaxy cluster}

This overdensity was first identified photometrically in COSMOS imaging using  Voronoi Tesselation \citep{Scoville2013}, which demonstrated a significant and concentrated surface overdensity of $>5$ galaxies per comoving Mpc$^{2}$ 
between $1.28<z<1.34$. A later analysis of deep XMM and Chandra X-ray data revealed extended emission indicating the presence of an ICM, with a well defined X-ray center \citep[][]{Goz2018, Goz2019}. The X-ray center is very close to a massive Log$_{10}$M$^*$/M$_{\odot}=11.2 $ quiescent galaxy targeted with ALMA spectroscopy (hereafter referred to as source 0) in \citet{Williams2021} at z=1.322. The group catalog published by \citet[][]{Goz2019} identified this overdensity as a cluster with  Log$_{10}$M$_{\rm halo}/$M$_{\odot}=13.6\pm0.1$, 
an estimated velocity dispersion of 360 km/s, and a virial radius of R$_{200}\sim$50 arcsec ($\sim$0.4 Mpc). \citet{Goz2019}  identified candidate cluster members from the COSMOS2015 photometric catalog \citep{Laigle2016}. A cluster redshift of $z=1.319$ was assigned based on the sole spectroscopic redshift among candidate cluster members \citep[][referred to herein as source 7]{Hasinger2018}. \citet[][]{Goz2019} label structures with Log$_{10}$M$_{\rm halo}/$M$_{\odot}<14$ as low mass clusters or groups. Given that the boundary between massive groups and low mass clusters is generally ill defined in the literature, in this work we consider this structure to be a low mass cluster, due to its extended X-ray emission and likelihood of hosting a massive brightest cluster galaxy (BCG; see sources 0 and 7 described below).

We note that given the modest significance \citep[3.3$\sigma$;][]{Goz2019} of the X-ray detection, hydrostatic equilibrium (i.e. virialization) was assumed in order to determine a halo mass from the X-ray emission using the low-scatter scaling relations for relaxed clusters \citep[e.g.][]{Kravtsov2006, Vikhlinin2009, Mantz2010, Leauthaud2010}. 
Significant deviations from hydrostatic equilibrium in clusters with an established ICM can result from merger events \citep[][]{RandallSarazinRicker2002, Poole2007, Wik2008}; however, these deviations cannot be distinguished in the existing data.  Therefore we also assume a virialized state applies for both the ICM gas and cluster members \citep[see][for a review]{Rosati2002}  throughout this work.
The location of the X-ray center, Voronoi-identified overdensity, candidate cluster members and our new spectroscopically confirmed galaxies (introduced in the next sections) are shown in Figure \ref{fig:image}.

\subsection{ALMA data}
The ALMA observations probing the CO overdensity were
carried out in project 2018.1.01739.S (PI: Williams) for target galaxy 34879 \citep[][ listed as source ID=0 in Table \ref{tab:sed}]{Williams2021}. The field was observed on 2018 December 18 and 2019 January 17 using the Band 3 (3~mm) receivers. The correlator was configured to center the CO(2–1)
line for 34879 at z=1.322 (99.284 GHz) within a spectral window of 1.875 GHz width, providing $\sim$5500 km s$^{-1}$ of bandwidth centered on the expected frequency of the CO line. Three additional spectral windows were used for continuum observations. The target was observed for a total of $\sim$3.2 hours on-source. The array was in a compact configuration
yielding a synthesized beam size of 2.7'' $\times$ 1.9''. Our data reduction procedure is outlined in \citet[][]{Williams2019, Williams2021}.  The spectral cube of our reduced data have a resulting noise of  $\approx$30$\mu$Jy/beam in a 400km/s channel measured near the rest-frequency of
the CO(2–1) line ($\nu_{rest}$=230.538 GHz). The 100\,GHz continuum data reach a sensitivity $\approx$5$\mu$Jy/beam at the phase center. None of our newly discovered CO-emitters are significantly detected in the continuum, though source IDs 1 and 5 are marginally detected at the 2$\lesssim$$\sigma$$\lesssim$3 level. These nondetections are unsurprising given the depth of our data and typical assumptions for the dust temperature, emissivity, and gas-to-dust mass ratio.

\begin{figure*}
\begin{center}
    \includegraphics[scale=.8, trim=0 0 20 0, clip]{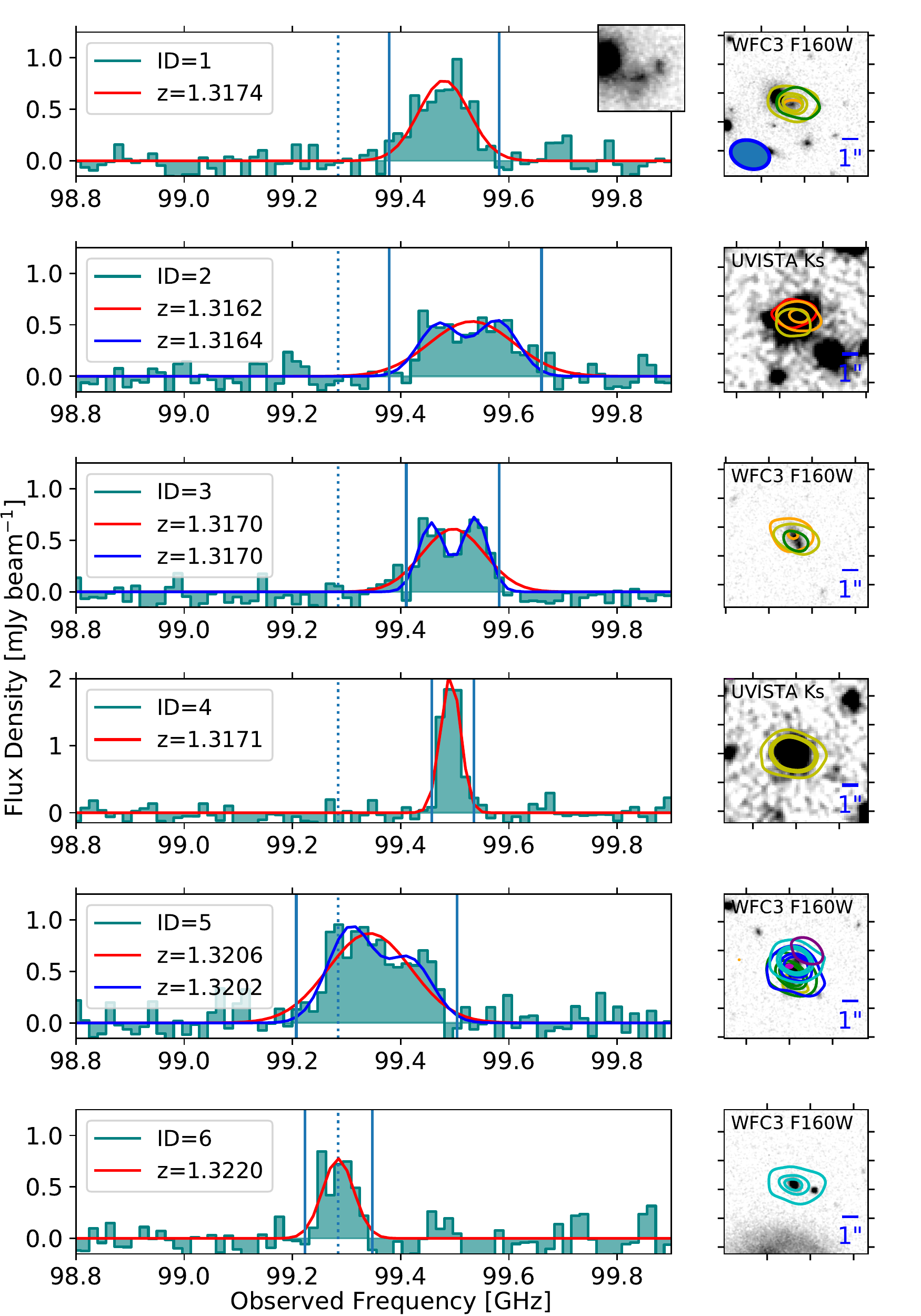}
    \caption{Left panel: our ALMA CO(2-1) spectra for the six emitters in 50 km/s channels. Dashed line corresponds to z=1.322, the redshift of the quiescent galaxy target from \citep{Williams2021}. Red lines are single Gaussian fits to the spectra and blue are double Gaussian (redshifts as measured from each in the legend). The  subpanel shown for ID=1 is a zoom-in of the WFC3/F160W image showing the  substructure that is blended in all catalogs (see Figure \ref{fig:radio}) with the primary quiescent galaxy (ID=0). Right panel: near-infrared image cutouts of each source, with ALMA contours overplotted. ALMA beam size is shown as the blue ellipse in top right panel. Contours are measured in 200 km/s channels and correspond to 25, 50, and 60 mJy/beam km/s levels. This cluster lies on the edge of existing HST/WFC3 F160W imaging from CANDELS.  UltraVISTA Ks band imaging is used for those galaxies without HST coverage.    }
    \label{fig:spec}
   \end{center}
\end{figure*}

We significantly detected 6 sources of CO(2-1) line emission (at $>5\sigma$, referenced herein as source IDs 1-6), easily identified through inspection of the CO spectral cube. To extract spectra for each source, we used
the \texttt{uvmultifit} package \citep[][]{MartiVidal2014} to
fit multiple pointlike sources to the visibility data. Briefly, we perform a joint fit of all sources at each frequency channel, using approximate by-eye source positions as the starting estimates, with the exception of the spectroscopically-identified but CO-undetected source ID 0, whose position is fixed to the phase center. 

To measure CO(2-1) line fluxes and luminosities, we first fit a simple Gaussian profile to the extracted spectrum. In a few cases (ALMA ID=2, 3, and 5) there is clear evidence for either gas rotation or multiple components in the CO(2-1) spectrum. For these sources, we also fit a double Gaussian model, where we fix the FWHM to be the same for both components, and define the redshift as the mean of the two peaks. Our results are not dependent on the single or double Gaussian assumption. We adopt the double Gaussian to model the line for those 3 sources. To measure the integrated CO(2-1) line flux, we integrate the observed spectrum over the frequency range where the Gaussian model is greater than 5\% the peak value and add in quadrature each channel rms to measure the integrated line flux uncertainty. Integrating the Gaussian model produces consistent integrated line flux measurements within the uncertainties, but our method more accurately captures the line flux in cases where the line profile is not perfectly described as a Gaussian.

Two existing optical spectroscopically-confirmed galaxies (undetected in CO) are associated with this overdensity: the galaxy associated with the group by \citet[][]{Goz2019} measured by \citet[][source 7]{Hasinger2018} and our target quiescent galaxy \citep[source 0;][]{Belli2014a,Williams2021}, in addition to our 6 new CO(2-1) sources.
Our measurement of upper limits for the two undetected galaxies and our procedure to convert from CO(2-1) line luminosity to M$_{\rm H_{2}}$ follow that outlined in \citet{Williams2021}. Briefly, we use a channel width of 500 km/s to measure upper limits to M$_{\rm H_{2}}$ for the two undetected sources with spectroscopic redshifts (IDs 0 and 7).

To convert CO(2–1) luminosity to a molecular gas mass, M$_{\rm H_{2}}$, we assume a luminosity ratio between the CO(2–1) and CO(1–0) transitions r$_{21}$ = 0.8 in temperature units and a CO-H$_2$ conversion factor $\alpha_{\rm co}$ = 4.4 M  (K km s$^{-1}$pc$^2$)$^{-1}$. Based on fundamental metallicity relations \citep[e.g.][]{Genzel2015}, our galaxies (with exception of ID=1) likely have high metallicity, close to solar (Log$_{10}$Z/Z$_{\odot}\sim$0.1). Models for the variation of $\alpha_{\rm co}$ at these high metallicities do not predict much variation and are consistent with Milky Way-like,  \citep[see e.g. Fig. 9 in][and references therein]{Bolatto2013}. This indicates that it is unlikely that $\alpha_{\rm co}$ could be higher, and therefore the Milky Way-like value is a reasonable, and even conservative assumption for our sample. Additionally, the presence of warm or high velocity dispersion gas would only serve to reduce $\alpha_{\rm co}$, and therefore the inferred gas masses, even further \citep[][]{Narayanan2012}. Our molecular gas properties and limits are listed in Table \ref{tab:mol} and their CO spectra and images are presented in Figure \ref{fig:spec}.

\begin{figure*}
    \includegraphics[scale=.6, trim=60 400 0 60, clip]{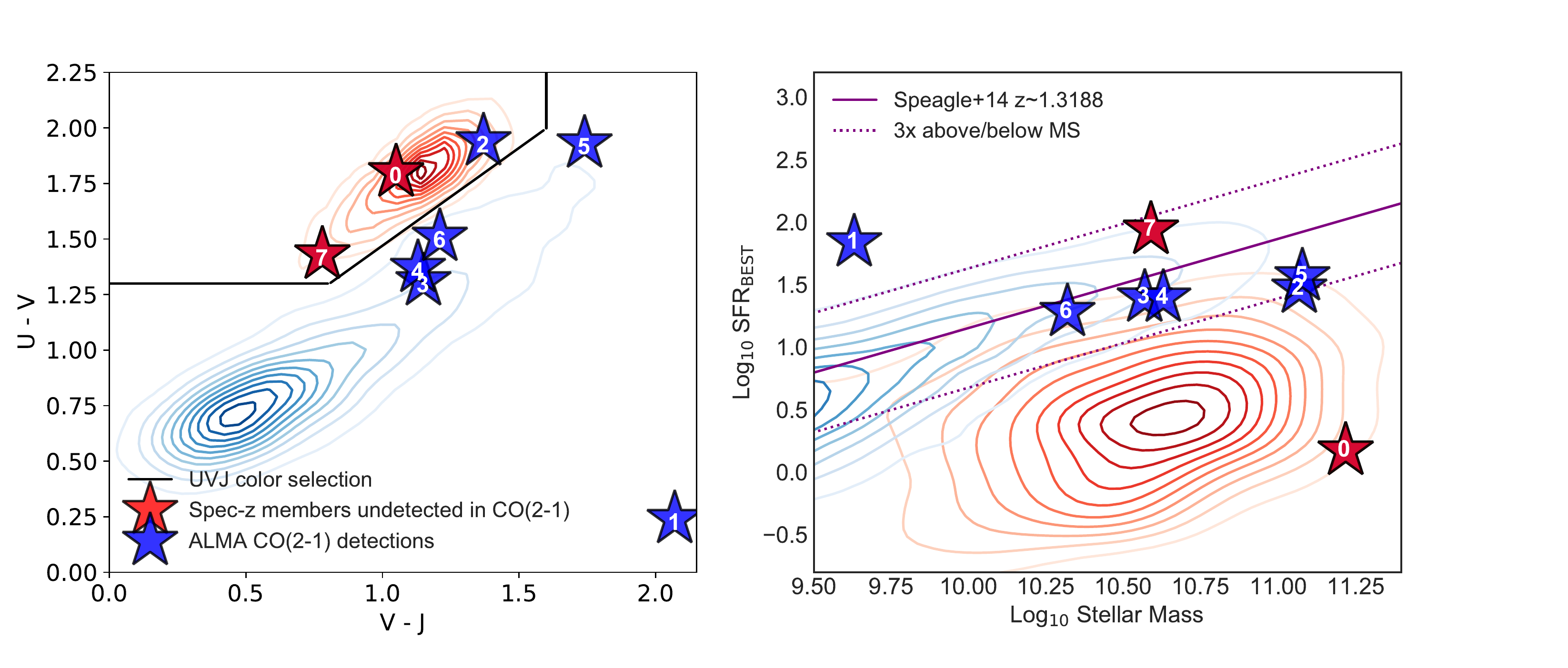}
    \caption{Properties measured by \texttt{BAGPIPES} for CO(2-1) emitters plus the two spectroscopically confirmed cluster members for which we have a CO(2-1) upper limit. The left panel shows restframe U-V vs V-J colors for star-forming and quiescent 3D-HST galaxies (blue and red contours, respectively) at $1<z<1.5$. UVJ colors for our spectroscopically confirmed cluster members are measured with \texttt{BAGPIPES}. Right panel: the star-formation-M* diagram, with the main sequence as measured by \citep[][]{Speagle2014}. For 3D-HST galaxies, we use SFR$_{\rm UV+IR}$. For our sample we use SFR$_{best}$ as defined as in Table \ref{tab:sed}. Points are labeled by their ALMA ID.}
    \label{fig:diag}
\end{figure*}

\subsection{Opt/IR data}\label{sec:optir}
We have a total of 8 spectroscopic cluster members, including the two galaxies with optical spectra previously measured (IDs 0, 7) and our 6 new CO(2-1) sources (IDs 1-6). This cluster lies in the gap between WFC3 and ACS grism data from the 3DHST program and therefore no grism redshifts are available. 

We identified counterparts to galaxies with CO(2-1) emission using the ULTRAVISTA photometry catalog \citep[][]{Muzzin2013}.  5 out of 6 CO sources have counterparts, with the exception of the source with ALMA ID=1. This source was previously noted in \citet{Williams2021} to be a companion of the target quiescent galaxy (ID=0) at a distance of $\sim1''$  and velocity offset of 600 km/s. This velocity offset is significantly larger than the expected spectroscopic redshift uncertainty, given that the measured velocity dispersion of the quiescent galaxy $\sigma=213\pm53$ km/s \citep[][]{Belli2014a}. 
This source is blended with the quiescent galaxy in all public photometric catalogs we explored \citep{Muzzin2013, Laigle2016, Skelton2014, Whitaker2011}, although the large velocity offset measured with our CO spectra suggest it is a separate system (although likely to merge in future). Moreover, the high-resolution HST data indicate that the companion source may itself be a blend of 3 clumps or different satellite galaxies (see inset in Figure \ref{fig:spec}). However, the components are unresolved at the low spatial resolution of our ALMA data and the CO line does not show evidence of multiple components.

We therefore extracted the flux density of ID=1 through the deblending code \texttt{Mophongo} (\citealt{labbe2006, labbe2010a, labbe2010b, labbe2013, labbe2015}). Briefly, this tool leverages a higher resolution map to reconstruct the brightness profiles, and remove the contribution, of all sources within a radius of $9\farcs0$ from the source of interest. Aperture photometry is then performed on the neighbour-clean stamp, and corrected to total using the brightness profile on the low-resolution image and the PSF reconstructed at the specific location of the target. For our analysis, we adopted the combined F125W, F140W and F160W mosaics from the 3D-HST program (\citealt{Skelton2014, Momcheva2016}) as high-resolution prior, and extracted the flux density using a $1\farcs8\ $-diameter aperture. Given the low spatial resolution of the CO map, we forced the extraction of the photometry for ID=1 by placing a synthetic point-source at the location corresponding to the peak of the CO emission. Specifically, we extracted the photometry in the CFHTLS \citep{Erben2009, Hildebrandt2009} $u*, g, r, i,$ and $z$ bands, Subaru/Suprime-Cam $Bj,Vj, g^+, r^+, i^+$ and $z^+$ \citep{Taniguchi2007}, Subaru HyperSuprimeCam $g, r, i, z$, and $y$ \citep{Aihara2018a,Aihara2018b,Aihara2019}, $Y, J, H$, and K$_S$ from the DR4 of the UltraVISTA program \citep{McCracken2012}, IRAC $3.6$ and $4.5\mu$m from S-CANDELS \citep{Ashby2015}, and IRAC $5.8$ and $8.0\mu$m from the S-COSMOS project \citep{Sanders2007}.

Given the lack of evidence of multiple independent galaxies, we assume all optical/infrared components make up one single companion. 
We note that the fainter two components display bluer colors than the brightest primary component of the companion, possibly caused by patches of unobscured star formation within the source. The brightest primary component is very red, and lacks any detectable flux blueward of WFC3/F160W. 

For results presented herein, we re-fit the UV to near-IR photometry from the UltraVISTA catalog for IDs 2-7, and our deblended photometry for IDs 0-1, uniformly using the SED-fitting code Bayesian Analysis of Galaxies for Physical
Inference and Parameter EStimation (\texttt{BAGPIPES}) \citep{Carnall2018}. \texttt{BAGPIPES}  assumes the stellar population synthesis
models of \citet{Bruzual2003} and implements nebular
emission lines following the methodology of \citet{Byler2017}
using the \texttt{CLOUDY} photoionization code 
\citet{Ferland2017}. We fit the photometry of our sample assuming a delayed $\tau$-model star formation history (SFH), and the \citet{Charlot2000} dust attenuation model, to measure the stellar mass, SFR integrated over the last 100 Myr (SFR$_{SED}$), and rest-frame U-V vs V-J colors, for classifying galaxies as star-forming or quiescent \citep{Williams2009,Muzzin2013MF}. The location of our spectroscopically confirmed cluster members in the UVJ diagram is shown in the left panel of Figure \ref{fig:diag}.

We note that the spatially non-uniform colors in the deblended companion source (ID=1) results in unusual integrated blue U-V and red V-J restframe colors. We tested SED-fitting the individual deblended clumps of ALMA ID=1 separately, finding that the blue clumps have $>$10x lower stellar mass than the primary red source alone. Since they contribute only a tiny fraction of the overall mass (and we derive its SFR from the radio, see Section \ref{sec:sfr}) our results do not depend on whether we assume the blue clumps belong to the primary source or not, besides producing the odd U-V vs V-J colors.  Given the strange colors of ID=1 we also explore a uniform prior range in metallicity, as well as fixing metallicity to solar. As this does not change our results (stellar masses are all consistent) and we present results with metallicity fixed at solar. The SED-fitting results used in this work are listed in Table \ref{tab:sed}.

\subsection{Star-formation rates}\label{sec:sfr}

\begin{figure}[th]
    \includegraphics[scale=.7, trim=5 0 0 0, clip]{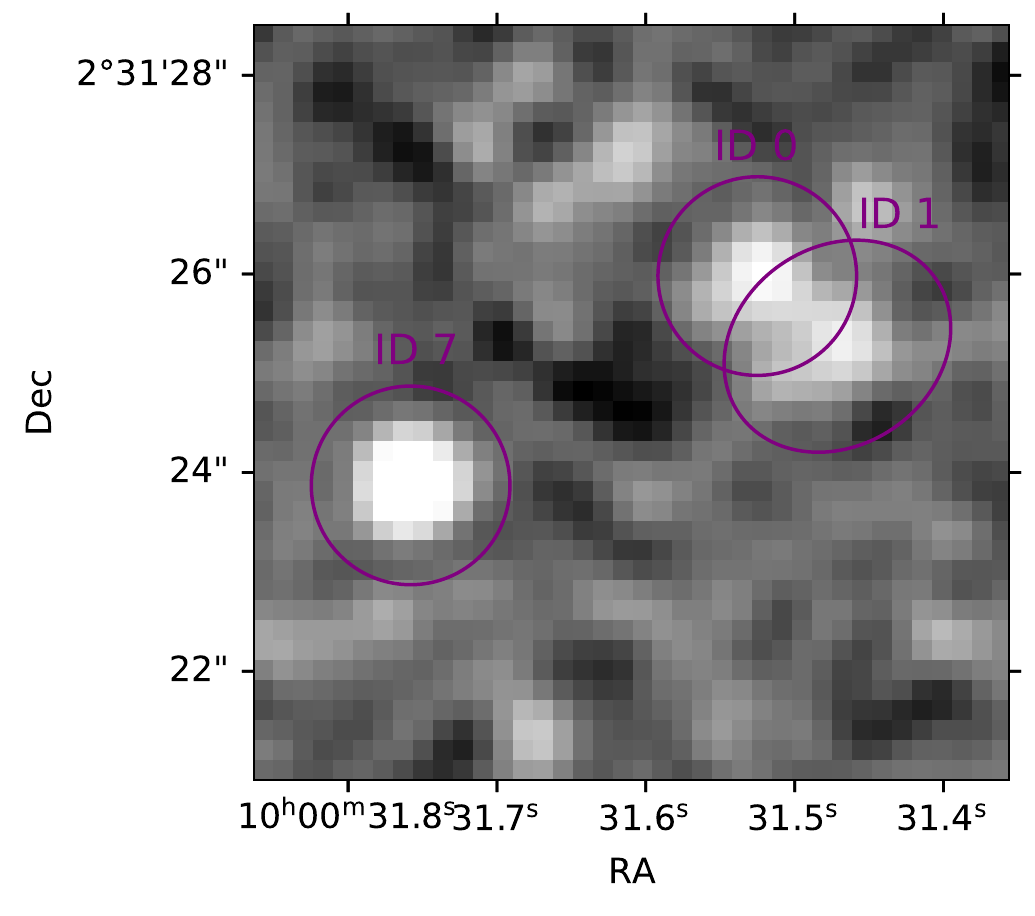}
   \caption{Radio image from the high resolution (0.75\arcsec) VLA-COSMOS 3 GHz Large Project \citep{Smolcic2017} demonstrating  IDs 0 and 1 are distinct radio sources, although blended in existing public catalogs. Our deblended radio fluxes for IDs 0 and 1 are presented in Section \ref{sec:sfr}. }
    \label{fig:radio}
\end{figure}

We use the deep multi-wavelength data from mid-infrared to radio in COSMOS to quantify and compare the SFRs for our CO-detected and non-detected sources. Unfortunately, given the proximity of  source 0 (the central quiescent galaxy) and source 1, we are unable to use any published SFR for these two sources, as they are blended in the ULTRAVISTA, 3DHST, and VLA 3 GHz radio catalogs \citep{Muzzin2013, Skelton2014, Smolcic2017, Algera2020}. However, our CO spectroscopy in combination with the rest-optical spectroscopy of \citep{Belli2014a} confirm that they are distinct sources (see dotted line in Figure \ref{fig:spec}). Though blended in available radio catalogs \citep{Smolcic2017, Algera2020}, we confirm visually that two distinct radio sources are identifiable with the 0.75\arcsec\ resolution from the VLA-COSMOS 3 GHz Large Project \citep{Smolcic2017}.  Leveraging our use of these spectroscopic priors, we perform source detection and photometry on the 3 GHz map using the {\tt Python Blob Detection and Source Finder (PyBDSF)} software package \citep[v1.9.2;][]{MohanRafferty2015}, modeling sources as Gaussians with conservative source detection parameters ({\tt thresh\_isl=3.0}, {\tt thresh\_pix=5.0}).  To deblend our sources, we turn off the grouping of nearby Gaussians into islands ({\tt group\_by\_isl=False}), which yields separate detections at moderate significance (see Figure \ref{fig:radio}).   We confirm that the deblended source fluxes total to the published blended flux within the uncertainties \citep[$38.6\pm2.9$ $\mu$Jy;][]{Smolcic2017}.

The quiescent galaxy (source 0) is both a radio source, deblended by our radio photometry, and a blended 24$\mu$m source.  Given its quiescent nature as confirmed by spectroscopy \citep{Belli2014a}, this likely indicates significant AGN activity, which cannot be disentangled from any SFR activity using the UV+IR SFR based on 24$\,\mu$m \citep[measured by][]{Muzzin2013} or the deblended radio flux.  Instead, in line with current philosophy in the literature that SED-fit based SFRs are less likely to be contaminated by AGN among quiescent galaxies \citep[e.g.][]{Fumagalli2014}, we use the SFR averaged over the last 100 Myr as measured with {\tt BAGPIPES} (see Section \ref{sec:optir}).

For source 1, we use the deblended 3 GHz flux and k-correct to 1.4 GHz assuming the convention $S_{\nu}\propto\nu^{\alpha}$ with $\alpha=-0.7$ \citep{Condon1992}. ALMA sources 2, 3, 4, and 5 are detected in the deeper radio imaging from \citet{Algera2020} and we use the k-corrected L$_{1.4GHz}$ from their public catalog to derive radio SFRs using the calibration 

\begin{equation}
    \log \left(\frac{\mathrm{SFR}}{M_{\odot} \mathrm{ yr}^{-1}}\right) = 0.823 \times \log \left(\frac{L_{1.4}}{\mathrm{W} \mathrm{ Hz}^{-1}}\right) - 17.5
\end{equation}

from \citet{Molnar2021}, which incorporates the luminosity dependence of the radio-infrared correlation.

Source 2 is undetected at $24\mu$m (used to measure UV+IR SFR) and therefore we adopt its radio-derived SFR of 29 M$_{\odot}$/yr, which is just below the sensitivity level of MIPS 24$\mu$m. However, we note that its SFR derived from SED-fitting to the optical/near-infrared data suggests a factor of $\sim3$ lower SFR, in line with its measured red UVJ colors. Since the source is not radio-loud, it is likely tracing the level of obscured SF missed by SED-fitting and not an AGN. 
For sources 3,4,5 the radio-derived SFR estimates are within a factor of 1.5x  those estimated based on UV+IR \citep[]{Muzzin2013}. To maximize consistency among our sample we adopt the radio-based SFR for sources 1-5 herein. 

Source 6 is undetected in any of the available radio imaging \citep[][]{Smolcic2017, Algera2020}, and is undetected at 24$\mu$m, making its UV+IR SFR based on this photometry unreliable \citep[][]{Muzzin2013}. For this source we again use the SFR averaged over the last 100 Myr from {\tt BAGPIPES}. 

Source 7 \citep[the spectroscopic source from][ without CO(2-1) emission]{Hasinger2018} sits squarely in IRAC AGN color space \citep{Donley2012, Kirkpatrick2013} 
and is also a radio and x-ray source. As such, its UV+IR SFR derived using MIPS is likely overestimated (it is 40$\times$ over the SED-fit SFR and 2.5$\times$ that from radio). However, the source is not radio-loud because it lies on the radio-infrared correlation \citep[][]{Helou1985, Condon1991, Yun2001} and thus the majority of radio flux most likely arises from star formation rather than the AGN \citep{Alberts2020}. For this source as well we thus adopt the radio-based SFR (87 M$_{\odot}$/yr) as with sources 1-5. We note that the SFR averaged over the last 100 Myr derived from SED-fitting suggests a lower SFR$\sim$5 M$_{\odot}$/yr, which is more in line with its red-UVJ colors near the post-starburst region of the diagram \citep{Belli2019}, but the source would remain within the scatter of the main sequence (see Figure \ref{fig:diag}). 

Our determined best estimates of the SFR as outlined above for each source (SFR$_{best}$) that is used in our analysis are quoted in Table \ref{tab:sed}. Based on these properties, 6 out of 8 of our sample would be classified as main sequence galaxies \citep[][see right panel of Figure~\ref{fig:diag}]{Speagle2014, Whitaker2014}. Source 0 is well below the main sequence and its companion source 1 lies above the main sequence.

\subsection{Summary of cluster members}\label{sec:clustsumm}

\begin{deluxetable*}{lccccccccc}[!th]
\tablecaption{Optical to radio SED properties } 
\tablecolumns{9}
\tablewidth{0pt}
\tablehead{
\colhead{ALMA ID} &
\colhead{ UltraVISTA ID} & 
\colhead{RA} & 
\colhead{Dec} & 
\colhead{Log$_{10}$M$^*$} & 
\colhead{SFR$_{SED}$ } & 
\colhead{SFR$_{UV+IR}$} & 
\colhead{SFR$_{radio}$} & 
\colhead{SFR$_{best}$ } 
}
\startdata
0 & 210589$^a$ & 150.131380 &  2.523800 & 11.22 & 1.47 & 22.90 & 57.23 & 1.47  \\
1 & -99 & 150.131012 &  2.523659 & 9.59 & 45.43 & -99.00 & 69.25 & 69.25  \\
2 & 210543 & 150.127225 &  2.523947 & 11.04 & 9.10 & 1.28 & 29.50 & 29.50  \\
3 & 210530 & 150.134492 &  2.524271 & 10.58 & 33.13 & 22.14 & 25.04 & 25.04  \\
4 & 210038 & 150.127542 &  2.518901 & 10.62 & 29.17 & 32.77 & 24.99 & 24.99  \\
5 & 209948 & 150.129054 &  2.516903 & 11.09 & 92.37 & 56.59 & 37.06 & 37.06  \\
6 & 210534 & 150.138625 &  2.525131 & 10.32 & 18.43 & 0.68 & -99.00 & 18.43  \\
7$^b$ &  210442 & 150.132310 & 2.523304 & 10.59 & 4.79 & 207.76 & 87.26 &  87.26  \\
\enddata 
\tablenotetext{a}{Referred to with ID=34879 in \citep[]{Belli2014a,Williams2021}}
\tablenotetext{b}{Spectroscopically confirmed member from \citep[][]{Hasinger2018}} 
\end{deluxetable*}\label{tab:sed}

\begin{deluxetable*}{lccccccccc}[!th]
\tablecaption{Molecular gas properties } 
\tablecolumns{12}
\tablewidth{0pt}
\tablehead{
\colhead{ALMA ID} &
\colhead{z$_{\rm co}$}  & 
\colhead{FWHM } &
\colhead{v$_{\rm offset}^b$ } &
\colhead{S$_{\nu}$d$\nu^c$ } &
\colhead{L'$_{\rm co}^c$} &
\colhead{M$_{\rm H_{2}}^c$}\\
\colhead{} & \colhead{} & 
\colhead{[km/s]} &
\colhead{[km/s]} &
\colhead{[mJy km/s]} &
\colhead{[10$^9$ K km/s pc$^2$ ]} &
\colhead{[10$^{10}$ M$_{\odot}$]}
}
\startdata
0$^a$ & 1.322 & - & 414 &  $<$13.8 &  $<$0.33  &  $<$0.55  \\
1  & 1.3174 & 331.6$\pm$30 & -177  &  283.1$\pm$21.1 & 6.46$\pm$0.5 & 3.55$\pm$0.3 \\
2  & 1.3164 & 553.8$\pm$56 & -312  &  295.1$\pm$20.4 & 6.72$\pm$0.5 & 3.70$\pm$0.3 \\
3  & 1.3170 & 411.2$\pm$42 & -229  &  242.9$\pm$15.2 & 5.54$\pm$0.3 & 3.05$\pm$0.2 \\
4  & 1.3171 & 129.4$\pm$8 & -214  &  279.3$\pm$14.2 & 6.37$\pm$0.3 & 3.50$\pm$0.2 \\
5  & 1.3202 & 544.7$\pm$49 & 179  &  472.5$\pm$30.4 & 10.83$\pm$0.7 & 5.96$\pm$0.4 \\
6  & 1.3220 & 205.5$\pm$35 & 413  &  159.6$\pm$20.9 & 3.67$\pm$0.5 & 2.02$\pm$0.3 \\
7$^{a}$ & 1.319 &  -  & 24 & $<$13.9 &  $<$0.33 &  $<$0.55 \\
\enddata 
\tablenotetext{a}{Undetected in ALMA data but with spectroscopic coverage of CO(2-1), limiting flux measured in a 500 km/s channel. We provide 1$\sigma$ upper limits for L'$_{\rm co}$ and 3$\sigma$ upper limits for M$_{\rm H_{2}}$ assuming R$_{21}$= 0.8 in temperature units, $\alpha_{\rm co}$= 4.4.  Molecular gas masses can be rescaled under different assumptions as M$_{\rm H_{2}}\times$(0.8/$_{21}$)($\alpha_{\rm co}$/4.4).}\label{tab:mol}
\tablenotetext{b}{Velocity offset relative to the cluster redshift ($z=3.188$) we calculate in Section \ref{sec:clustsumm}}
\tablenotetext{c}{ Integrating observed spectrum and propagating channel errors}
\end{deluxetable*}

14 photometric candidate group members were identified by \citet{Goz2019} ranging from $1.29<z_{phot}<1.35$. 9 of those 14 are within the 20\% ALMA primary beam area of our observations. Our target quiescent galaxy (source 0) spectroscopically confirmed by \citet{Belli2014a} was not identified as a group member because their spectroscopic redshifts were not included in the public COSMOS catalogs. With our new data, we spectroscopically confirm 3 photometric cluster members identified by \citet{Goz2019} (the other 6 that are within our primary beam are undetected, including source 7).  The other 3 CO sources were not identified as cluster members using the \citet[][]{Laigle2016} catalog but photometric redshifts from \citet{Muzzin2013} place the galaxies within $\delta z$ +/- 0.1 of the cluster redshift.
As source 1 was absent from optical/infrared and radio catalogs in COSMOS, it was not identified as a cluster member in the \citet{Goz2019} catalog. 8 spectroscopic members are now confirmed, all sources have spectroscopic redshifts that are within a range of $<$725 km/s. 

Given our new spectroscopic confirmations, we use these redshifts to re-calculate the cluster redshift and velocity dispersion using the bi-weight location method described in \citet{Beers1990} using the astropy stats function {\tt biweight\underline\ location} \citep{Astropy}. 
We  find a systemic redshift of  $z$=1.3188 (with a dispersion of $dz$= 0.00245 or 317 km/s) essentially the same as that in the published catalog (based only on the source 7 redshift of $z=1.319$).  Following \citep{Danese1980}, we correct for the contribution to the velocity dispersion introduced by the measurement uncertainty (assuming the typical redshift uncertainty of our ALMA data, 0.0001) and find a corrected value of $\sigma_v = $ 295 (+128,-55) km/s where these are the +/- 68\% uncertainties. This is consistent with that derived using the X-ray (360 km/s), and therefore adopt the X-ray dispersion in the next section.

\begin{figure*}
    \includegraphics[scale=.9, trim=10 0 0 0, clip]{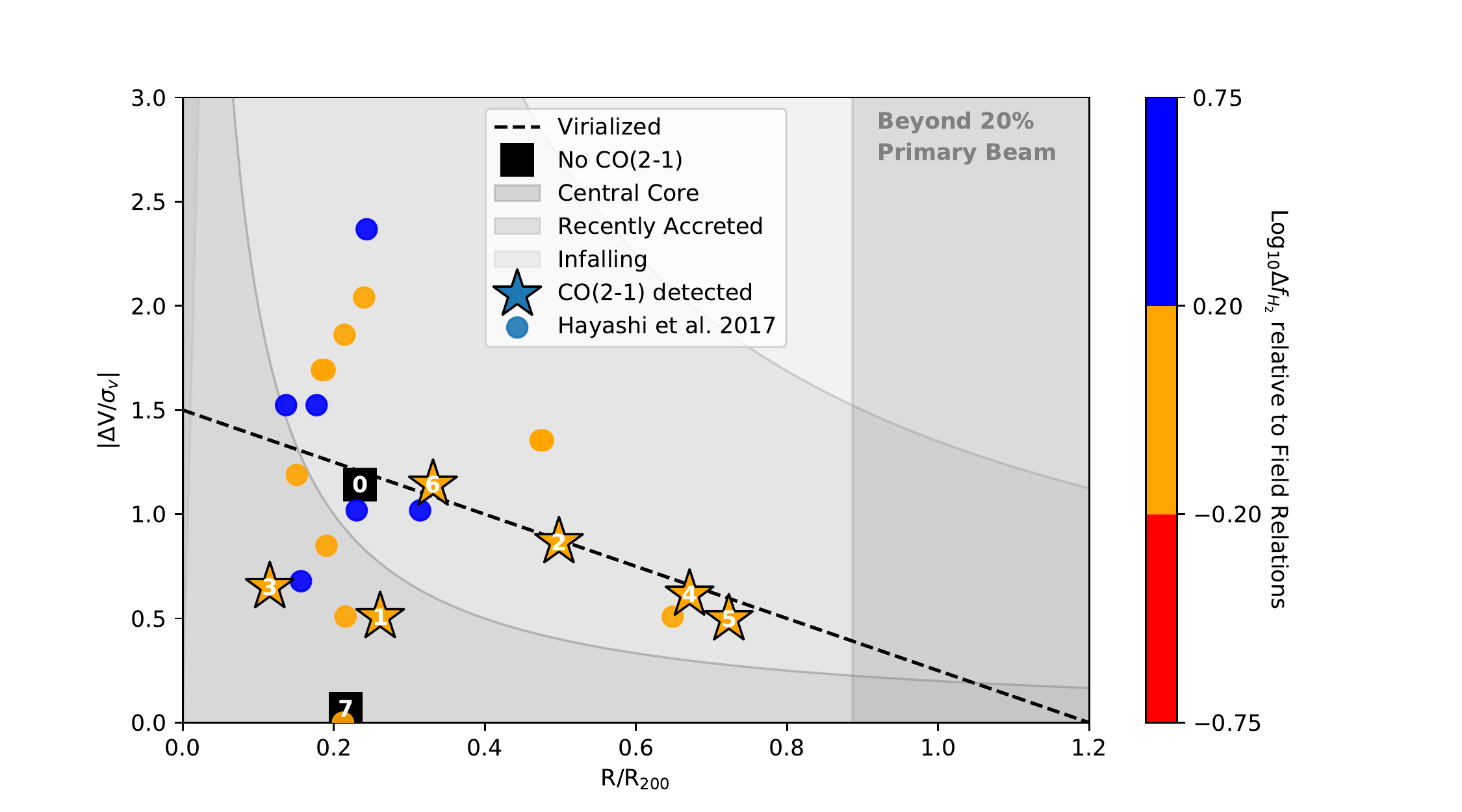}
    \caption{Phase diagram indicating accretion history of cluster galaxies. Galaxies are classified by their relative velocity to the cluster redshift (scaled by velocity dispersion of cluster) vs their distance from the X-ray-derived cluster center, normalized by virial radius. Shaded curves delineate regions as defined as in \citet{Noble2016,Hayashi2017} for central core, recently accreted (defined here to include their intermediate region) and infall region. All spectroscopic sources are within the virial radius R$_{200}$ = 50.8 arcsec = 425 kpc and are likely virialized. Area of phase space not probed by our data (beyond 20\% of the primary beam limit; $\sim$45 arcsec; M$_{\rm H_{2}}$ $>$ 4.2x 10$^{10}$M$_{\odot}$ at 5$\sigma$) is indicated by the gray box (R/R200 $> 0.9$). Our spectral coverage  ($\sim$ 5500 km/s bandwidth) limits lie off the plot at -5 and +7 on the y-axis. 
    For comparison, we include comparable data from a virialized, coeval but more massive (z=1.46, Log$_{10}$M$_{\rm halo}/$M$_{\odot}\sim14$ cluster  \citep[][]{Hayashi2017, Hayashi2018}. Both datasets indicate that gas survives in galaxies after first infall, and can even maintain field-like gas reservoirs within the core or virialized region.     }
    \label{fig:phasediag}
\end{figure*}

\section{Results} \label{sec:results}

\begin{figure}
    \centering
    \includegraphics[scale=0.4]{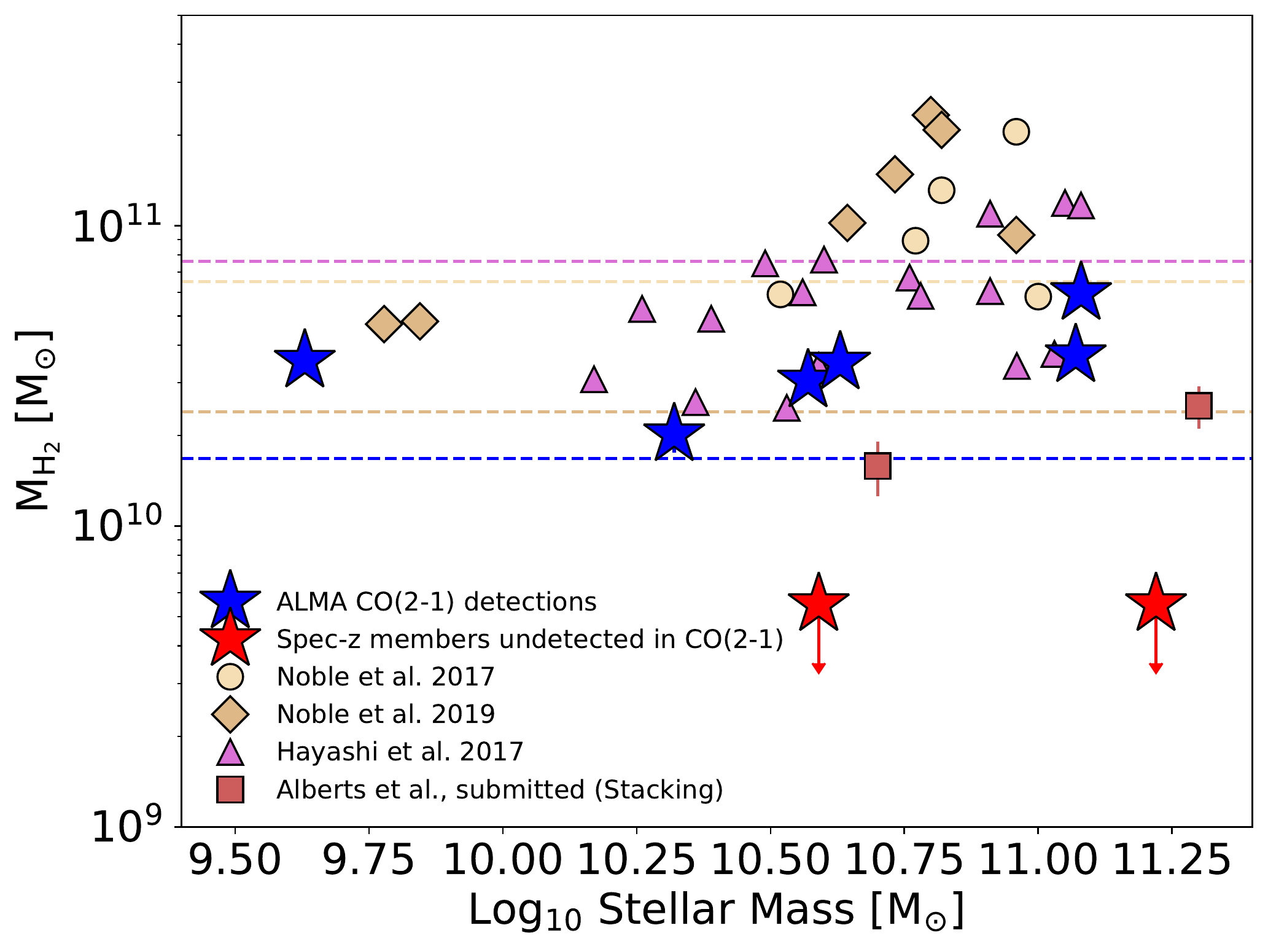}
    \caption{The molecular gas mass as a function of stellar mass for cluster studies at $z\sim1-2$.  Our CO(2-1) detected cluster members are shown as blue stars.  Spectroscopic, but CO-undetected members are shown as upper limits in red stars.  Cluster galaxies detected in CO at $z\sim1.5$ are shown in purple triangles \citep{Hayashi2017, Hayashi2018} and at $z\sim1.6$ in light yellow circles \citep{Noble2017} and dark yellow diamonds \citep{Noble2019}. The $5\sigma$ detection limit at $50\%$ of the ALMA beam is shown in dashed lines with corresponding colors; \citet{Hayashi2017, Hayashi2018, Noble2017} have limits roughly 4.5$\times$ shallower than this work, while \citet{Noble2019} presented follow-up at similar depth.  We note that the \citet{Hayashi2017} ALMA mosaic includes overlapping pointings, resulting in small areas deeper than the representative limit.  A dust continuum stacking study (Alberts et al., submitted) is shown for contrast in brown squares, representing the average gas masses of undetected $z\sim1-2$ star-forming cluster galaxies.}
    \label{fig:gas_masses}
\end{figure}

\begin{figure*}
    \centering
    \includegraphics[scale=0.37, trim=0 0 0 0, clip]{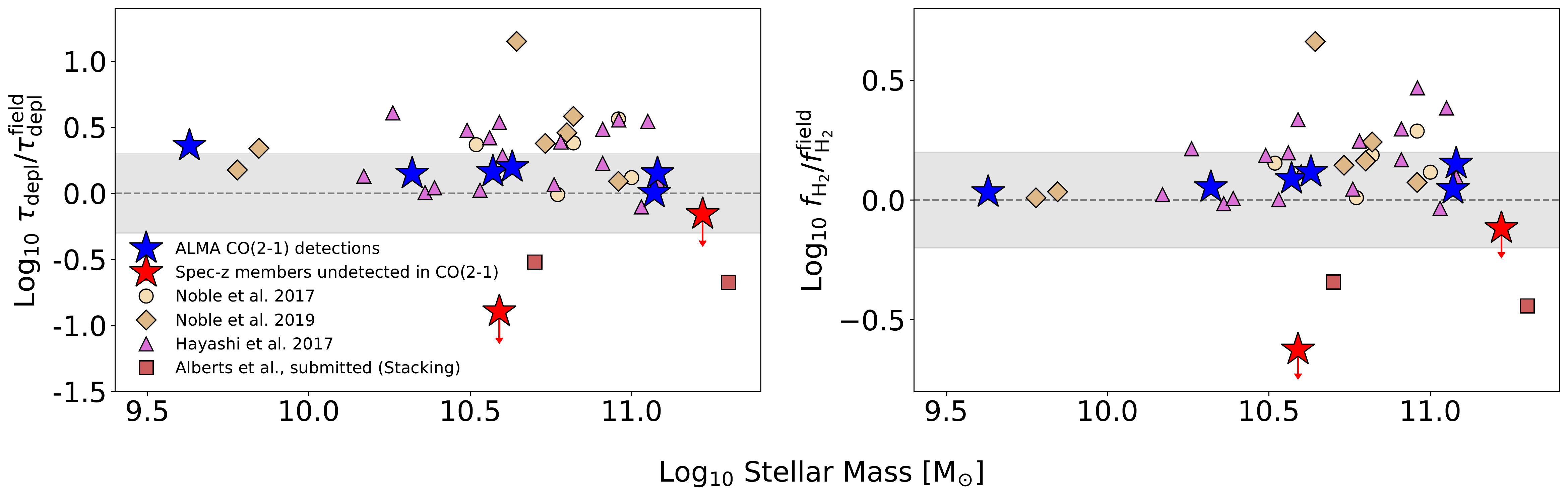}
    \caption{ The gas depletion timescales (left) and gas fractions (right) of cluster galaxies relative to the predicted field gas properties as a function of stellar mass.  The predicted field properties are derived for each individual source given its redshift, stellar mass and distance from the \citet{Speagle2014} main sequence.  Symbols are as in Figure~\ref{fig:gas_masses}.  The dashed line represents the \citet{Tacconi2018} field scaling relation for $\tau_{\rm depl}$ and $f_{\rm H_{2}}$ with an assumed scatter of 0.3 dex \citep{Liu2019} for the former and 0.2 dex for the latter (shaded regions).  As shown in Figure~\ref{fig:gas_masses}, our survey probes the entire main sequence at Log$_{10}$M$^*/M_{\odot}>10.5$, one of the deepest limits to date. The CO detected cluster members in this work are consistent with field-like gas properties, while other CO studies \citep{Hayashi2017, Hayashi2018, Noble2017, Noble2019} show a mix of galaxies on and above the field relations. Dust continuum stacking of star forming galaxies (brown squares; Alberts et al. submitted), on the other hand, places the average gas depletion timescales and gas fractions at shorter and lower than the field scaling relations. }

    \label{fig:scalrel}
\end{figure*}

\subsection{Accretion histories of cluster members}

This overdensity of CO emitters are all found in a relatively compact projected area on the sky. All 8 spectroscopic members lie within the projected virial radius (R$_{200}\sim$50 arcsec; $\sim$400 kpc). To explore their spatial distribution within the cluster and estimate the cluster accretion histories for each galaxy, we produce a phase-space diagram showing the velocities of the 8 cluster members relative to the cluster redshift ($\Delta$V, normalized by cluster velocity dispersion, $\sigma_{v}$, which was measured by \citet{Goz2019} using scaling relations for Lx-M$_{200}$ and Lx-$\sigma$; \cite[e.g.][]{Leauthaud2010, Connelly2012}) vs the galaxy projected distance from the X-ray-derived cluster center (normalized by virial radius; R/R$_{200}$). Simulations indicate that using projected distances and line-of-sight velocities instead of full 3D space still results in accurate characterization in phase space \citep[e.g.][]{Rhee2017}. 

The phase space diagram is shown in Figure \ref{fig:phasediag}, where the curves delineate the approximate regions where galaxies would be considered in the central core of the cluster, recently accreted, or still infalling. We delineate regions according to (R/R$_{200})\times(\Delta v/\sigma_v) < 0.2$ (cluster center), $0.2<($R/R$_{200})\times(\Delta v/\sigma_v) < 1.35$ (intermediate and recently accreted regions) and (R/R$_{200})\times(\Delta v/\sigma_v) > 1.35$, beyond which galaxies are still in their initial infall. These definitions of phase space regions are as outlined in both \citet{Noble2016} and \citet{Hayashi2017}. We additionally plot  a representation of the virialized region as defined in \citet[][]{Mahajan2011,Jaffe2015}. However, we note that these regions should be regarded as approximate since regions defined in projected phase space can contain interlopers \citep{Rhee2017}.  Additionally, 
\citet{Goz2019} report relatively low S/N on the measured X-ray luminosity and halo mass for this cluster, 3.3 and 5.4 respectively, which corresponds to low S/N on the derived virial radius and velocity dispersion.  However, we note that the X-ray derived values agree well with our spectroscopically derived cluster properties, as described in Section~\ref{sec:clustsumm}.

Given the wide bandwidth of the CO spectroscopy ($\sim$5500 km/s) we are sensitive to the overwhelming majority of possible velocities within the cluster center ($-7< \Delta V/\sigma_{v}<5$). Given our deep integration and the compact nature of this cluster,  we probe 90\% of the area within the virial radius ($R/R_{200}<0.9$) within 20\% of the ALMA primary beam to a 5$\sigma$ limit of M$_{\rm H_{2}}$ $>$ 4.2$\times10^{10}$ M$_{\odot}$. While we do not detect any conclusively infalling CO sources, 
our data does not probe a substantial part of the first infall region, which can stretch 2-3 times the size of the virial radius \citep[e.g.][]{Rhee2017, Zinger2018}.

As mentioned previously, caution is advised when interpreting phase space diagrams, since the correspondence between the denoted regions and physical accretion histories is approximate, with interlopers  possible in each given space \citep{Rhee2017}.  This makes the interpretation of small samples such as ours difficult.  As such, we focus our comparison  with other samples in the literature in order to place it in the context of these earlier CO studies. Relatively few clusters currently have molecular gas constraints at $z\sim1-2$ \citep{Noble2017, Noble2019, Rudnick2017}, and in addition, well measured cluster parameters \citep{Hayashi2017, Hayashi2018}. \cite{Hayashi2017, Hayashi2018} presented the largest sample of molecular gas in phase space in a coeval ($z=1.46$) but more massive X-ray cluster.  In general, we find that our cluster galaxies occupy the same phase space as the Hayashi et al. sample, mostly populating the recently accreted/intermediate region with some galaxies in the cluster core (Figure \ref{fig:phasediag}).  Our relative distribution between these two regions agrees with the conclusion from Hayashi et al. that fewer CO detections are found in the cluster cores compared to the intermediate/recently accreted region, indicating a relative depletion of gas in the core. However, with only 2 other spectroscopically confirmed galaxies in our cluster that lack CO detection, we are not spectroscopically complete enough to estimate the fraction of cluster members that have already lost the majority of gas. We note the caveat that the core region is defined relative to the X-ray center and assuming the cluster does not have significant sub-structure or asymmetries.

\subsection{Comparison to gas properties of field galaxies}

We expand this comparison by looking at the gas content in these cluster galaxies relative to that expected in coeval field galaxies.  In Figure \ref{fig:phasediag}, we color-code galaxies in the phase space diagram using their deviation from the molecular gas fraction $f_{\rm H_{2}}\equiv \frac{M_{\rm H_{2}}}{(M_{\rm H_{2}}+M^*)}$ predicted by the field scaling relations presented in \citet{Tacconi2018}  such that 
\begin{equation} 
\Delta f_{\rm H_{2}} = \frac{f_{\rm H_{2},\rm measured}}{<f_{\rm H_{2},\rm field}(z,\Delta MS,M^*)>}. 
\end{equation}

where $\Delta$MS is the deviation (in log space) from the star forming main sequence as defined by \citet{Speagle2014}. We bin the colorbar such that orange ($-0.2<$ Log$_{10}\Delta f_{\rm H_{2}}<0.2$ dex) corresponds to the range of systematic uncertainties in the molecular gas measurements that were used to build the scaling relation \citep[][]{Tacconi2020}. We consider this range in orange to be consistent with the field-calibrated scaling, Log$_{10}\Delta f_{\rm H_{2}}=0$ (i.e. field-like). Although we note that \citep[][]{Tacconi2018,Tacconi2020} do not characterize or quantify the scatter of the data used to measure the scaling relation, \citet{Liu2019} confirm that a range of  $0.15-0.25$ dex conservatively represents the typical systematic uncertainties and is also comparable to the scatter of molecular gas measurements at these redshifts.

We also color code the \citet{Hayashi2018} points by their deviation from the field main sequence relations. Their work used a metallicity dependent $\alpha_{\rm CO}$ conversion factor \citep{Genzel2012}. To ensure a fair comparison to our work, we re-calculate M$_{\rm H_{2}}$ from their measured L$^{\prime}_{\rm CO}$ using our assumed $\alpha_{\rm CO}=4.4$, consistent with the value commonly adopted for field scaling relations \citep[][]{Tacconi2020}.  
Using the recalculated values, we then measure $f_{\rm H_{2}}$ and compare to the field scaling relations.  We find that the majority (65\%) of the \cite{Hayashi2017,Hayashi2018} CO-detected galaxies are consistent with field-like gas content, and 35\% of CO-detected galaxies are in excess of field expectations (i.e. gas enhanced).

In contrast, all of our CO detections have gas fractions that are  consistent within the scatter of field-based scaling relations at similar SFR and M* (median Log$_{10}\Delta f_{\rm H_{2}}=0.05$ dex). Of the two CO undetected galaxies in the cluster core, limits suggest very low gas content, consistent with their UVJ quiescent status.
We additionally note that our conclusions do not change had we adopted a metallicity dependent $\alpha_{\rm co}$ for our lowest-mass (and therefore presumably lowest metallicity) source ID=1, whose inferred $f_{\rm H_{2}}$ would still be consistent with the field relations.

We next expand our comparison\footnote{Molecular gas constraints also exist for galaxies in overdensities or protoclusters at $z>2$ \citep[][]{Umehata2017, Dannerbauer2017, Lee2017, Zavala2019, Wang2018, Castignani2019, GomezGuijarro2019,Long2020, Champagne2021, Hill2021, Shen2021, Jin2021}, however given that protoclusters are likely to be characterized by significantly different dynamical states than a relaxed, virialized cluster,  we do not directly compare to them.} to include the sample of CO detections in $z\sim1.6$ cluster galaxies presented in \citet{Noble2017, Noble2019}.  As their molecular gas measurements were made with similar assumptions as this work, no modifications are made.   In Figure~\ref{fig:gas_masses}, we present the H$_2$ masses based on CO detections for co-eval cluster galaxies at $z\sim1-2$, along with the detection limits (5$\sigma$ at $50\%$ of the ALMA primary beam) for each work (M$_{\rm H_{2}}^{\rm lim}\sim6.5\times10^{10}$ and 7.6$\times10^{10}$ M$_{\odot}$ for \cite{Noble2017} and \cite{Hayashi2018}, respectively).  Our 5$\sigma$ detection limit at $50\%$ the primary beam (1.6$\times10^{10}$M$_{\odot}$) is nominally 4-4.5 times deeper than \citet{Noble2017, Hayashi2017, Hayashi2018}, though overlap in multiple pointings in the latter survey probe lower $M_{\rm H_{2}}$ in small areas.  \citet{Noble2019} presented data of similar depth to this work (M$_{\rm H_{2}}^{\rm lim}\sim1.95\times10^{10}$M$_{\odot}$) over a single pointing, detecting 4 new galaxies as compared to the \citet{Noble2017} observations.

Overall, the CO detections in this work sit at relatively lower gas masses compared to previous works.  Our data further puts strong upper limits on our two CO-undetected cluster members, with gas masses less than 5.5$\times 10^9$M$_{\odot}$.  Incomplete spectroscopy in the optical/infrared, however, prevents us from ruling out more gas-poor cluster members.  To explore this issue, we compare to a recent study of molecular gas using stacking of dust continuum measurements in 11 massive clusters at $z\sim1-2$ (Alberts et al. submitted).  In field galaxies, CO and dust continuum yield comparable measurements of the molecular gas \citep{Tacconi2020}; differences may exist in overdense environments, but this issue has only been explored in small samples of proto-cluster galaxies \citep[e.g.][]{Lee2021}. For now, we assume they are comparable within the uncertainties of this study. In the mass bin most comparable to this work (Log$_{10}$M$^*$/M$_{\odot}\sim10.75$), Alberts et al. found that the average molecular gas in cluster galaxies sits at 1.6$\times10^{10}$ M$_{\odot}$, right at our $5\sigma$ detection limit.  At higher masses, the average gas mass may be above our detection limit; however, the small area covered may be insufficient to observe these rarer more massive galaxies.  This comparison suggests that gas-poor cluster members may still be missing from studies of detected galaxies, even given the deep detection limits presented in this work. 

To put the gas properties of these $z\sim1-2$ cluster galaxies in the context of the field, we derive the field-relative gas depletion timescales and gas fractions (Figure~\ref{fig:scalrel}) by dividing by the predicted values via the \citet{Tacconi2018} scaling relation, as described for Figure \ref{fig:phasediag}.  The predicted field values are calculated using the redshift, stellar mass, and distance from the main sequence ($\Delta$MS) of each cluster galaxy, providing a direct comparison.  
The scatter in $\tau_{\rm depl}/\tau_{\rm depl}^{\rm field}$ is taken to be 0.3 dex as in \citet{Liu2019}, incorporating measurement and systematic uncertainties in the gas masses and measured SFRs. The scatter in $f_{\rm H_{2}}/f_{\rm H_{2}}^{\rm field}$ is adopted as 0.2 dex, assuming that the scatter in the ratio is dominated by the uncertainties in the gas masses rather than by systematic errors in the stellar mass measurements.  Our deep detection limit enables us to probe the entire main sequence at Log$_{10}$M$^*/$M$_{\odot}>10.5$ down to gas masses of M$_{\rm H_{2}}\gtrsim1.6\times$10$^{10}$M$_{\odot}$, from which we can safely rule out the presence of cluster galaxies with extended gas depletion timescales or enhanced gas fractions, out to R/R$_{200}>0.2$ and $\Delta v/\sigma_v>1.5$.  This is in contrast to existing CO surveys which find enhanced gas content (e.g. M$_{\rm H_{2}}>6-8\times10^{10}$M$_{\odot}$), with the corresponding potential for long depletion timescales, modulo uncertainties in SFR measurements, and enhanced gas fractions in excess of the field.  These gas enhanced members are even found in the cluster cores and the virialized regions \citep[][see Figure \ref{fig:phasediag}]{Hayashi2017, Hayashi2018}.  On the other hand, the stacked average gas properties of star forming cluster members at $z\sim1-1.75$ (Alberts et al. submitted), probing below current detection limits, suggest shorter depletion timescales and lower gas fractions, below the field scaling relations at fixed stellar mass and distance from the main sequence, are still missing from these analyses. We discuss this possibility further in Section \ref{sec:discussion}.

Finally, we note that both sources 0 and 7, which lie in the quiescent region of UVJ color space and are likely in the cluster core, have low limits on any molecular gas fractions ($<$3 and $<$12\%, respectively), in line with accumulating evidence that quiescent galaxies have cold molecular gas reservoirs of order a few percent or less \citep{Bezanson2019, Williams2021, Caliendo2021, Whitaker2021}. However, both of these sources are detected at 24$\mu$m and 3 GHz. They are similar to the sample studied by \citet{Belli2021} in that they were selected for quiescent optical/near-infrared spectral energy distribution but relatively bright emission at 24 $\mu$m. By contrast, \citet[][]{Belli2021} finds diverse gas reservoirs that are substantially higher $f_{\rm H_{2}}$ (13-23\%) for similar IR-bright quiescent galaxies (although they do not have IRAC colors reflective of AGN, such as our source 7). The number of such sources with measured gas reservoirs is small and thus we cannot reach definitive conclusions, but it is interesting that our two UVJ-quiescent, non-detected cluster members exhibit substantially lower gas reservoirs in contrast to \citet{Belli2021}.

\section{Discussion and Conclusions} \label{sec:discussion}

In this work, we have presented a serendipitous discovery of CO emitters in the central core and recently accreted regions of phase space in a low mass (Log$_{10}$M$_{\rm halo}$/M$_{\odot} =13.6$) cluster at $z=1.3188$.  This ultra-deep ALMA data provides evidence that molecular gas survives first infall into the cluster environment, and even maintains field-like gas content in main sequence galaxies (for all 6 CO-detected sources). An additional two members spectroscopically confirmed via optical/infrared spectroscopy 
are not detected in CO(2-1) indicating gas loss consistent with the evidence that they are in a more evolved, quiescent state.

Despite this deep data, however, we do not find evidence for enhanced gas content within $\sim$90\% of the virial radius. This result differs from comparable CO studies of coeval clusters at $z\sim1-2$. Where gas is detected, these studies mostly find that cluster members have gas reservoirs consistent with the field \citep[e.g.][]{Rudnick2017}, but some have also found a significant fraction of members hosting
elevated gas content relative to the field scaling relations \citep[in particular][see Figure~\ref{fig:scalrel}]{Noble2017, Noble2019, Hayashi2017, Hayashi2018}. There are a few things to consider in interpreting this.  First, it is important to note the differences in the relative sizes of the areas probed.
Our survey is the smallest among \citealt{Hayashi2017, Noble2017}, with only 1 ALMA pointing (45" radius at 20\% the primary beam). Comparatively, \citet{Noble2017} probe nearly 4$\times$ this area, and \citet{Hayashi2017} nearly 1.5$\times$, which may have increased the chances to catch rarer gas enhanced sources.

Second, we point out that given that \citet{Noble2017, Noble2019, Rudnick2017, Hayashi2017, Hayashi2018} are some of the first constraints on CO in high-redshift cluster galaxies, the targets were chosen to be highly star forming, and therefore likely containing large gas reservoirs, in order to maximize detection of CO rather than to create an unbiased sample.  As a serendipitous detection, our cluster is not biased in this way, however, even combined these cluster studies are still in the regime of small number statistics.

Given the caveats above, are these galaxies with enhanced gas content rare and unique among the more field-like cluster members? Interestingly, even those galaxies consistent with the field scaling relations remain within the upper envelope of the scatter.
Given the relatively high detection limit of the earlier surveys, 
it is plausible that main sequence populations of galaxies with lower gas content were simply missed. 
However, an interesting case is the deeper ALMA followup of one cluster in \citet[][]{Noble2017} published by \citet{Noble2019}. While \citealt{Noble2019} achieved a depth more comparable to our data for the one cluster (see Figure \ref{fig:gas_masses}), even this study found that the new CO detections among previously undetected lower mass objects remain in the upper envelope of the scatter of the field scaling relations, in good agreement with the sources presented in this work. The reason for the perceived dearth of gas reservoirs below the field scaling relation (encompassing the lower envelope of the scatter), 
is not clear (see Figure \ref{fig:scalrel}), though we note that the deep \citet{Noble2019} coverage is currently only of a single $\sim$48" radius pointing (at 20\% of the primary beam), and does not provide a uniform mapping over the entire cluster.
In summary, from the small number of clusters surveyed, there appears to be some tentative evidence for systematically larger molecular gas reservoirs relative to the field in CO detected sources (although see \citealt{Coogan2018} for one counter example at higher redshift), even if the systematic increase largely remains within the overall scatter of field galaxies. This may hint at a shared mechanism for the cluster members displaying enhanced gas fractions.

Wider surveys in a statistical sample of clusters to the depths presented here still might not be enough to present the full picture of molecular gas in cluster galaxies. There is emerging evidence from stacking of dust continuum emission that the cold gas reservoirs of star forming cluster galaxies is, on average, depressed below the field scaling relations at z$\sim$0.7 \citep[][]{Betti2019}, $z\sim1-2$ (Figure \ref{fig:scalrel}; Alberts et al. submitted) and in $z\sim$2 protocluster galaxies \citep{Zavala2019}.  Unfortunately this regime has not fully been reached by current CO surveys; the Alberts et al. results place the average gas mass at our $5\sigma$ detection limit.  This further highlights the mystery of CO detections inhabiting the upper envelope of the field scaling relations: if the average gas mass is somewhere below the field, in a regime where gas has been lost relative to comparable field galaxies, where are the gas-deficient ``sub-field" cluster members?  Understanding the distribution of gas masses among cluster galaxies is key to constraining the mechanisms responsible for gas enhancement or gas loss, and will require detections probing below the predicted field levels of gas content over larger areas of the cluster environment.  The latter is motivated by the wide area covered by Alberts et al., which found sub-field gas content out to 2R$_{vir}$ over 11 similarly selected clusters (Figure \ref{fig:scalrel}), more effectively mitigating the impact of cluster to cluster variation and the potential biases in observing only a small portion the cluster environment.

Alongside additional observations, the large range in gas properties suggested by detection and stacking studies, plus the aforementioned cluster-to-cluster variation, stress the importance of simulations in interpreting small empirical samples.  Recently, cosmological simulations have begun to address the question of gas properties in cluster environments.  They find significant gas loss staring at large radii ($>2$R$_{\rm vir}$) ranging from the stripping of hot halo gas \citep{Zinger2018} to the complete removal of all molecular gas \citep{Arthur2019, Mostoghiu2021} by the first passage of the cluster core \citep[see also][]{OmanHudson2016, Oman2021}.  In the first scenario, tightly bound disk gas is retained enough for star formation to proceed unimpeded for a time, consistent with the ``delayed, then quenched" scenario proposed by \citet{Wetzel2013} and supported by subsequent studies \citep{Haines2015, BaheMccarthy2015,OmanHudson2016,Maier2019,Rhee2020,Cortese2021}.  Depending on the fraction of total gas removed by stripping, our field-like gas masses may be consistent with the removal of hot halo gas, while more extreme stripping may be supported by the stacking results.  Populations with enhanced gas content are not well represented in these simulations; however, simulations have explored the idea of gas streams penetrating the cluster ICM \citep{Zinger2016}, which may provide a mechanism for delivering new gas to cluster galaxies such as has been plausibly observed at lower redshifts in BCGs \citep{Castignani2020c, Dunne2021}.

Our new data makes this cluster the lowest mass cluster at $z>1$ with both evidence of virialization and characterization of the molecular gas reservoirs.  Thus, this study bridges the observational gap between gas properties of the most massive clusters during this epoch, and those in the field.  More systematic surveys of molecular gas using deep CO or dust emission of clusters across halo mass are needed to constrain the physical processes impacting star formation activity during the critical cluster transition era of $1<z<2$.

\acknowledgements
We thank Ghassem Gozaliasl and Alexis Finoguenov for sharing their COSMOS group catalog, and Ivo Labb\'{e} for allowing use of the \texttt{Mophongo} software.
CCW acknowledges support from  NIRCam Development Contract NAS5-02105 from NASA Goddard Space Flight Center to the University of Arizona.
JS acknowledges support provided by NASA through NASA Hubble Fellowship grant \#HF2-51446 awarded by the Space Telescope Science Institute, which is operated by the Association of Universities for Research in Astronomy, Inc., for NASA, under contract NAS5-26555.  KEW acknowledges funding from the Alfred P. Sloan Foundation grant No. FG-2019-12514.
This paper makes use of the following ALMA data: ADS/JAO.ALMA \#2018.1.01739.S. ALMA is a partnership of ESO (representing its member states), NSF (USA), and NINS (Japan), together with NRC (Canada), NSC and ASIAA (Taiwan), and KASI (Republic of Korea), in cooperation with the Republic of Chile. The Joint ALMA Observatory is operated by ESO, AUI/NRAO, and NAOJ. The National Radio Astronomy Observatory is a facility of the National Science Foundation operated under cooperative agreement by Associated Universities, Inc. The Cosmic Dawn Center is funded by the Danish National Research Foundation.
\\
\\
Software: NumPy \citep{har20},
Matplotlib \citep{hun07}, 
Astropy \citep{astropy:2018}, 
SciPy \citep{mckinney-proc-scipy-2010}, 
Seaborn \citep{Waskom2021}, 
CASA \citep{mcm07}

\bibliography{manu}{}
\bibliographystyle{aasjournal}

\end{document}